\documentclass[graybox]{svmult}

\usepackage{url}
\usepackage{type1cm}        
\usepackage{float}                            
\usepackage{makeidx}         
\usepackage{graphicx}        
\usepackage{multicol}        
\usepackage[bottom]{footmisc}

\usepackage{newtxtext}       %
\usepackage{newtxmath}       

\makeindex             

\begin{document}

\title*{A transition of shared mobility in metro cities– a challenge post Covid-19 lockdown }
\author{Mohd Aman and Bushra Miftah}
\institute{Mohd Aman \at Department of Applied Mathematics, Faculty of Engineering and Technology, Aligarh Muslim University, Aligarh 202002, India, \email{mohdaman.maths@gmail.com}
\and Bushra Miftah \at Department of English, Faculty of Arts, Aligarh Muslim University, Aligarh 202002, India, \email{bushramiftah.in@gmail.com}}
%
%
\maketitle


\abstract{This chapter is written for the welfare of the society, questioning and enlightening the effects of the increment or decrement in the percentage of quality of air causing pollution due to the rise in the traffic post-lockdown due to COVID-19 in metro cities, specifically in Delhi. In this chapter, we address the question about people's preference in moving in the shared taxis to their workplaces or their reluctance and denial of the idea of moving in the shared vehicle because of the fear of getting infected. The sensitivity of the situation will compel the people to move in a single occupied vehicle (SOV). The rise in the number of vehicles on the roads will result in traffic jams and different kinds of pollution where people battling with the pandemic will inevitably get exposed to other health-related issues. We use a BPR (Bureau of Public Roads) model to combat this issue endangering the environment and public health. We exploit the BPR function to relate average travel time to the estimated number of commuters travelling by car. We collect mode share data from the NITI Ayog, the State Resource Centre and other authentic sources, which gives unique figures of the impact of shared mobility in India and how, in its absence, various sectors will get affected. Using the given data and the BPR, we evaluate increased vehicle volumes on the road if diﬀerent portions of transit and carpool users switch to single-occupancy vehicles and its effect on multiple other factors. Based on the study of densely populated city, Delhi, we predict that cities with significant transit ridership are at risk for extreme traffic and pollution unless transit systems can resume safe with effective protocols.}

\section{Introduction}
\label{sec:1}
\noindent The $21$st century saw the emergence of a newly found coronavirus, also known as the SARS (Severe Acute Respiratory Syndrome) COV--2 in the Wuhan city of China. Unaware of the cause of transmission to the human body, the deadly virus which had been declared Public Health Emergency of International Concern by WHO earlier this year has been contracted by more than one crore population across the globe causing more than 1 lakh fatalities bringing the entire world to a standstill.  

India, which holds a population of more than 130 crores has been severely affected by the newly found virus. However, to curb the same India went under what is known as the strictest lockdown on March 22, 2020. The stay-at-home orders had a drastic effect on the life of every common man. It caused a direct impact on the usage of the transport system. As the stay-at-home order took effect, ridership data show steep decline in both transit ridership and vehicular traffic.
 
\begin{figure}[H]
	\centering
	\sidecaption
	\includegraphics[scale=.25]{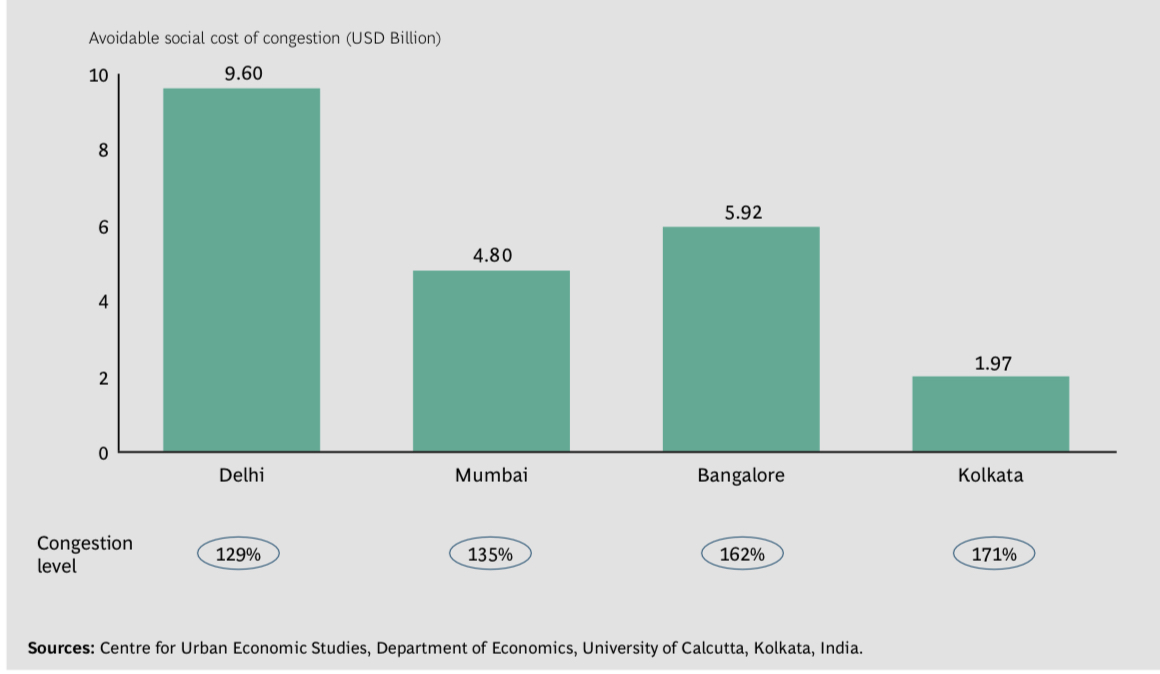}
	%
	%
	\caption{Cost of congestion across cities (2017) in USD Billion}
	\label{fig:a}       
\end{figure}

\begin{figure}[H]
	\centering
	\sidecaption
	\includegraphics[scale=.25]{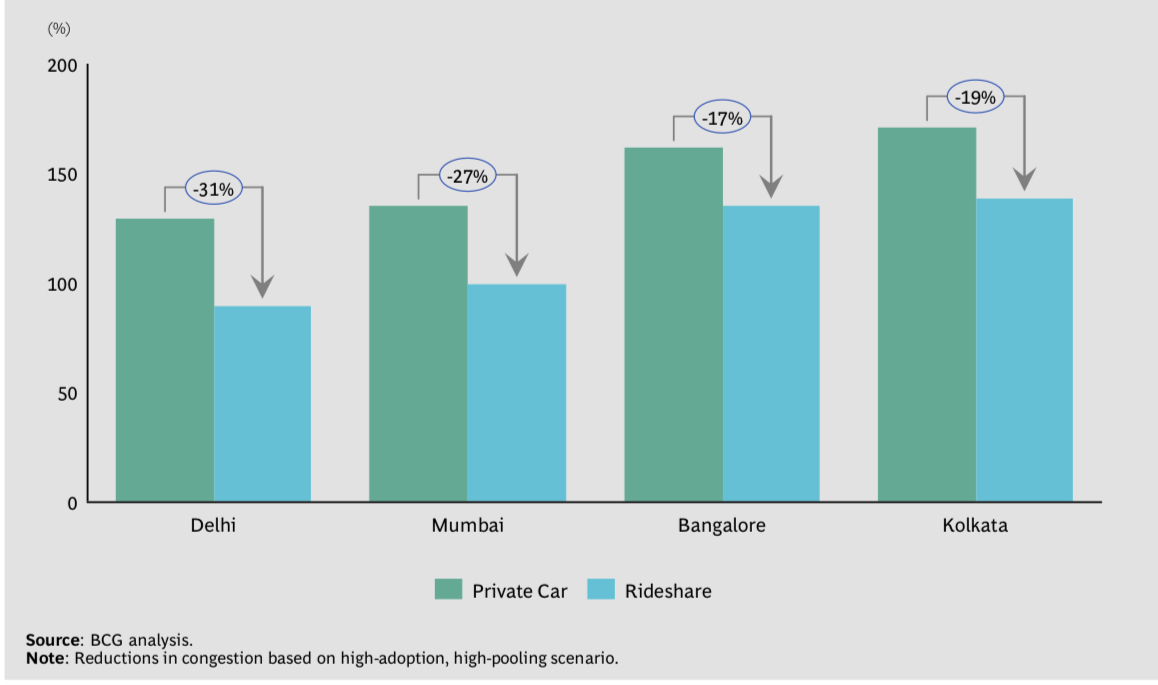}
	%
	%
	\caption{Road congestion during peak hours before vs. after rideshare (2017)}
	\label{fig:b}       
\end{figure}

According to the data collected in from NITI Ayog \cite{niti,niti2}, a State-Resource-Centre of Government of India established with the aim to achieve sustainable development goals in the economic policy-making process; there was an usurp growth in the vehicle ownership and transportation in the last few years which led to overcrowding and unsorted situation that resulted in the loss of $1.42$ lakh crore INR annually alone in the cities of Delhi, Mumbai, Kolkata and Bangalore. It has had severe and adverse implications on the health and environment. The increment in the number of vehicles had severely affected India's energy consumption, energy security and economy, pollution, congestion, health and safety. The graphs in Fig. \ref{fig:a} and \ref{fig:b} depicts a more detailed picture of the congestion and its cost across metro cities.


The transportation sector in India accounts for $18\%$ of commercial energy consumption and is highly dependent on oil imports. India imported $80\%$ of its oil at the cost of Rs. $4.2$ lakh crore in recent years. Additionally, private vehicle use has significant implications on land requirement for parking: in Delhi, for example, parking accounts for 8--10\% of the available land pool.

\begin{figure}[b]
	\centering
	\sidecaption
	\includegraphics[scale=.25]{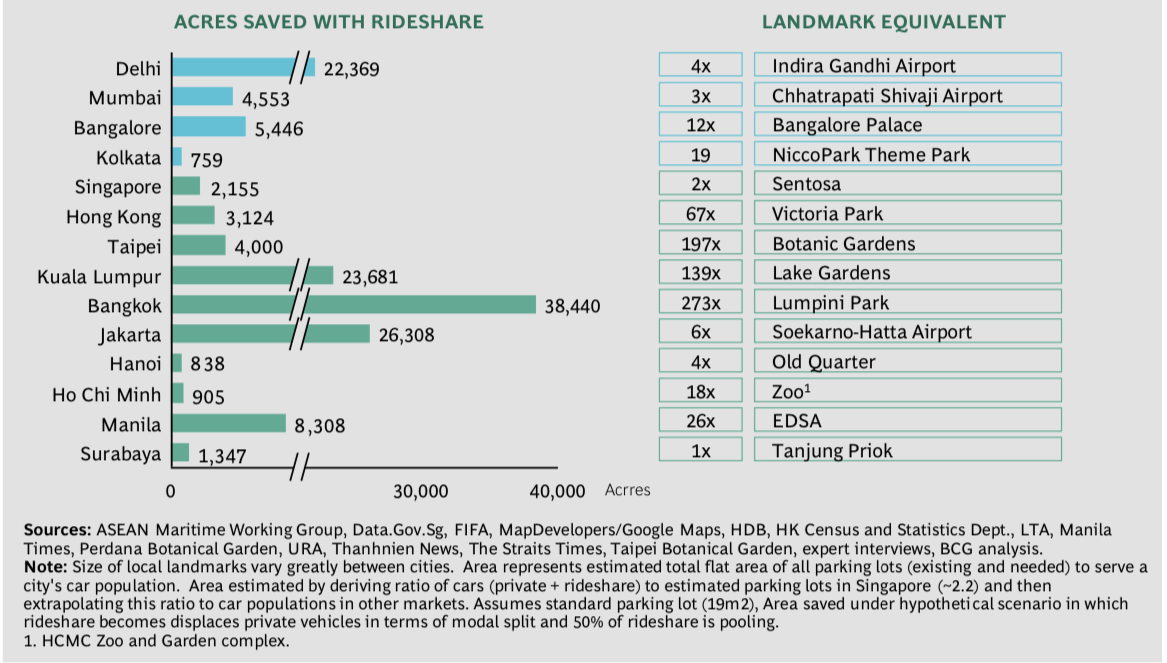}
	%
	%
	\caption{Estimated space that can be saved by adopting rideshare assuming rideshare substitutes for private cars (2017)}
	\label{fig:c}       
\end{figure}

\vspace{.35cm}
There is a direct effect of ridesharing practice on space. The figure \ref{fig:c} gives the idea of estimated space that could be saved by switching to rideshare from private vehicles in big cities in 2017.

Weighing the seriousness of the situation, it is deducted that with the complete unlocking of the lockdown, there will be a drastic mode shift of transportation to SOV (Single Occupied Vehicles).

Definitely, there will be a massive change in the traffic volume caused due to a sudden drop in transit usage and its impact on the commute time on the road.  With the collected data from verified sources, we estimate the number of cars on the road if certain proportions of commuters switch from transit and carpool modes to Single Occupancy Vehicle (SOV), thus increasing the total number of vehicles on the road and the average travel time. The total traffic volume is therefore taken as the number of vehicles commuting from this mode. This data when applied on the BPR model \cite{BPR,assignment,daskin,modeshift} to calculate the traffic volume and average traffic time give us an idea of the travel time in a condition when there is a decrease or the increase in the number of vehicles on the road.

Several researchers have carried out studies trying to model the congested traffic in the urban network for different countries across the globe. For e.g., Saberi et al. \cite{saberi} use a susceptible-Infected-recovered (SIR) model to describe the spreading of traffic jam. Colak et al. \cite{cloak} use the ratio of road supply to travel demand as a dimensionless factor to explain the percentage of time lost in congestion. The macroscopic BPR model is studied by Wong et al. \cite{wong} and Kucharski and Drabicki \cite{rafal} to relate traffic demand and travel time. One or several cities or regions are studied in the above works. Our research is specific to a densely populated Indian metro city, Delhi with an exorbitant increase in the traffic. There is no work done using the BPR analytical model on Indian metro cities which makes an inquiry with regards to road management in Indian cities and make predictions for their travel time under various scenarios of possible traffic volume.
\begin{figure}[b]
	\centering
	\sidecaption
	\includegraphics[scale=.65]{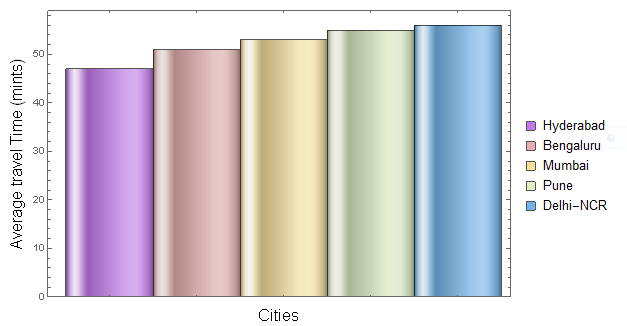}
	%
	%
	\caption{Travel Time of congested metro cities of India (2018)}
	\label{fig:t-t}       
\end{figure}
India accounts for top $15$ most polluted cities in the world. The figure \ref{fig:t-t} gives the average travel time of four highly congested metro cities in India in 2018.

The major finding on the collected data throws light on the greater risk of high transit cities of increased traffic so that system can resume safely without any complications.

The high transit metro cities are at a higher risk with potentially 4--30 minutes additional commute travel time in a one-way trip. This unusual rise in road transport which will result in a sudden increment in the number of vehicles will cause serious health and environmental issues. This can be curbed, and the transit ridership can be brought to normalcy with returns in trips by car, thereby bringing down the traffic.

\vspace{.25cm}
The rest of the chapter deals with the introduction of the formulation of BPR model, the introduction of data on traffic in metro cities of India, and show the forecast result of traffic in different reopening scenarios.
 
Our main ﬁnding is that the high--transit cities are at greater risk for increased traffic volume unless their transit systems can resume safe, high throughput operations quickly. 

\section{BPR model}
\label{sec:2}


%
\noindent A widely used impedance calculation model, the BPR (Bureau of Public Roads) model \cite{BPR,assignment,daskin,modeshift} states that traffic flow affects the traffic speed. The Bureau of Public Roads is a federal agency created in 1918. Lodged in the US Department, it has taken up much responsibility of building roads in national parks and forests, assisted states with road construction, helped beautify highways, and conducted various transportation studies. 

The (Bureau of Public Roads) BPR function \cite{BPR,assignment} is a model that relates the volume of traffic on the road to the travel time to traverse it. It has a great deal of usage in transportation management \cite{highway,capacity} and network traffic stimulation \cite{florian}. According to recent studies, apart from being applied on single--road traffic \cite{wong,rafal}, it also has applicability on urban scale transportation analysis. A BPR function thus provides us with a theoretical foundation of predicting metro area travel congestion based on traffic volume. 

The relationship between travel time and the volume (total no. of vehicles) can be expressed by the classical BPR function in the following way:
\begin{equation}\label{eq:01}
\mathcal{T}= \mathcal{T}_{0}\left[1+a\left(\frac{N}{C}\right)^{b}\right],
\end{equation}
where,
$\mathcal{T}$ is the travel time as a function of $N$ the volume on the roadway. The parameters $\mathcal{T}_{0}$, $C$ are the free ﬂow travel time and the practical road capacity respectively. The shape parameters a and b can be ﬁt or assumed to follow a common choice of $a=0.15$ and $b=4$.

In order to apply the function at the scale of a city, we interpret the BPR model as a function that takes the total number of road users in a city $N$ as a proxy for volume and returns the average commute travel time $\mathcal{T}$ in the city. In this interpretation, the capacity $C$ should be viewed as the total number of road users that can be accommodated in the city before average travel time begins to increase. The average travel time is overall users at all time of the day.

Rewriting \eqref{eq:01}, we have
\begin{equation}\label{eq:02}
\mathcal{T}= \mathcal{T}_{0}\left[1+0.15\left(\frac{N}{C}\right)^{b}\right],
\end{equation}
or
\begin{equation}\label{eq:03}
\mathcal{T}= \mathcal{T}_{0}\,\mu+N.
\end{equation}
The above equation clearly shows that the travel time $\mathcal{T}$ and the traffic volume $N$ have a linear relationship. The model parameters in \eqref{eq:01} can be estimated using linear regression. With the historic data from NITI Ayog and other sources \cite{niti,niti2,bandyo,shaikh,bcg}, we conduct a linear regresson on the data from 2012--2018 to find the coefficients $\mathcal{T}_{0}$ and $\mu$. This allows us to estimate the free-flow travel time $\mathcal{T}_{0}$ and the practical road capacity
\begin{equation}
c=\left(\frac{0.15\,\mathcal{T}}{\mu}\right),
\end{equation} 
for the metro city Delhi deriving from \eqref{eq:02}.
\section{Data Analysis \& Implementation}
\label{sec:3}
\subsection{Data Description}
\noindent The data used in the given equation has been derived mainly from NITI Ayog which is a State-of-Art Resource Centre with in-demand knowledge and some other sources as well \cite{niti,niti2,bandyo,shaikh,bcg}. Using the data, we can answer the following questions:
\begin{verbatim}
How many cars are on the road in each city?
\end{verbatim}
\begin{verbatim}
What is the average travel time experienced by commuters in each city?
\end{verbatim}


The data records means of transportation to work by selected characteristics. For Delhi, annual statistics of a given time (2012--2018) is recorded including the total number of road users for each commute mode (shared vehicles including public transit, single-occupancy vehicles (SOV)), as well as the average commute time.

\begin{figure}[H]
		\centering
	\sidecaption
	\includegraphics[scale=.55]{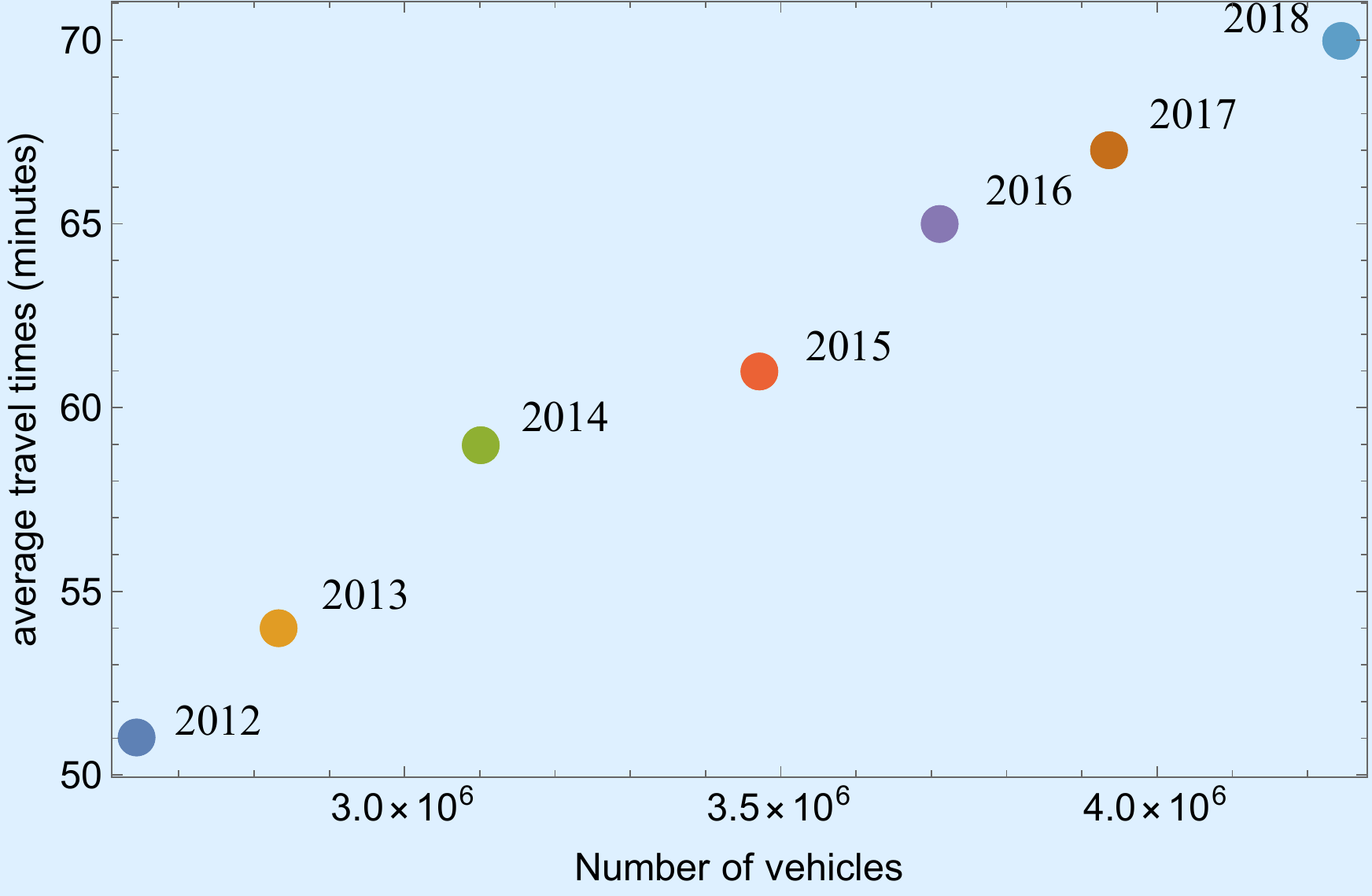}
	%
	%
	\caption{Observed data of number of vehicles and average one way tavel time (Delhi) from 2012--2018}
	\label{fig:t-n}       
\end{figure}

The road commute metrics are calculated with the help of raw data. The average travel time of the four-wheelers is taken. The total number of vehicles is computed by converting cars, jeeps, shared vehicles and buses into equivalent vehicles on the road. 

We have estimated the resulting vehicles on the road when $25\%$, $50\%$, $75\%$ and $100\%$ of commuters using shared vehicles switch to SOV. 

We use the latest statistics from the $2018$ data to estimate the number of vehicles on the road. The data from 2018 to present can be considered in an extension and is not taken into account in this analysis. We also note that the mode share data for taxis and ride-hailing is mixed in with motorcycles, bike and others and we omit the inﬂuence of taxis (not OLA, UBER, etc.) in this study, and do not count taxis in the total number of vehicles. The table \ref{tab:1} gives the historical commute data of Delhi from 2012 to 2018.

\begin{table}[H]
	\centering
	\caption{Exploitation of Historical Data$^a$ \& Data computed using BPR function (Delhi)}
	\label{tab:1}       
	%
	%
	\begin{tabular}{p{2cm}p{2.4cm}p{2cm}p{4.9cm}p{1cm}}
		\hline\noalign{\smallskip}
		Year & No. of Vehicles & Avg. Travel & Trave Time Ratio ~~  Capacity Ratio\\& &  Time  \\
		\noalign{\smallskip}\svhline\noalign{\smallskip}
		2012 & 2644520  & 51 & 1.007 ~~~~~~~~~~~~~~~~~~~~~~~ 0.807\\
		2013 & 2831460 & 54 & 1.067 ~~~~~~~~~~~~~~~~~~~~~~~ 0.864\\
		2014 & 3102442 & 59 & 1.165 ~~~~~~~~~~~~~~~~~~~~~~~ 0.946\\
		2015 & 3470526  & 61 & 1.205 ~~~~~~~~~~~~~~~~~~~~~~~ 1.059\\
		2016 & 3711060  & 65 & 1.284 ~~~~~~~~~~~~~~~~~~~~~~~ 1.132\\
		2017 & 3937104  & 67 & 1.324 ~~~~~~~~~~~~~~~~~~~~~~~ 1.201\\
		2018 & 4245760  & 70 & 1.383 ~~~~~~~~~~~~~~~~~~~~~~~ 1.295\\
		\noalign{\smallskip}\hline\noalign{\smallskip}
	\end{tabular}
	$^a$ Historical data collected from NITI Ayog \& other reliable sources
\end{table}


\subsection{Model Application \& Results }
\subsection*{BPR Model}
\begin{figure}[H]
	\centering
	\sidecaption
	\includegraphics[scale=.55]{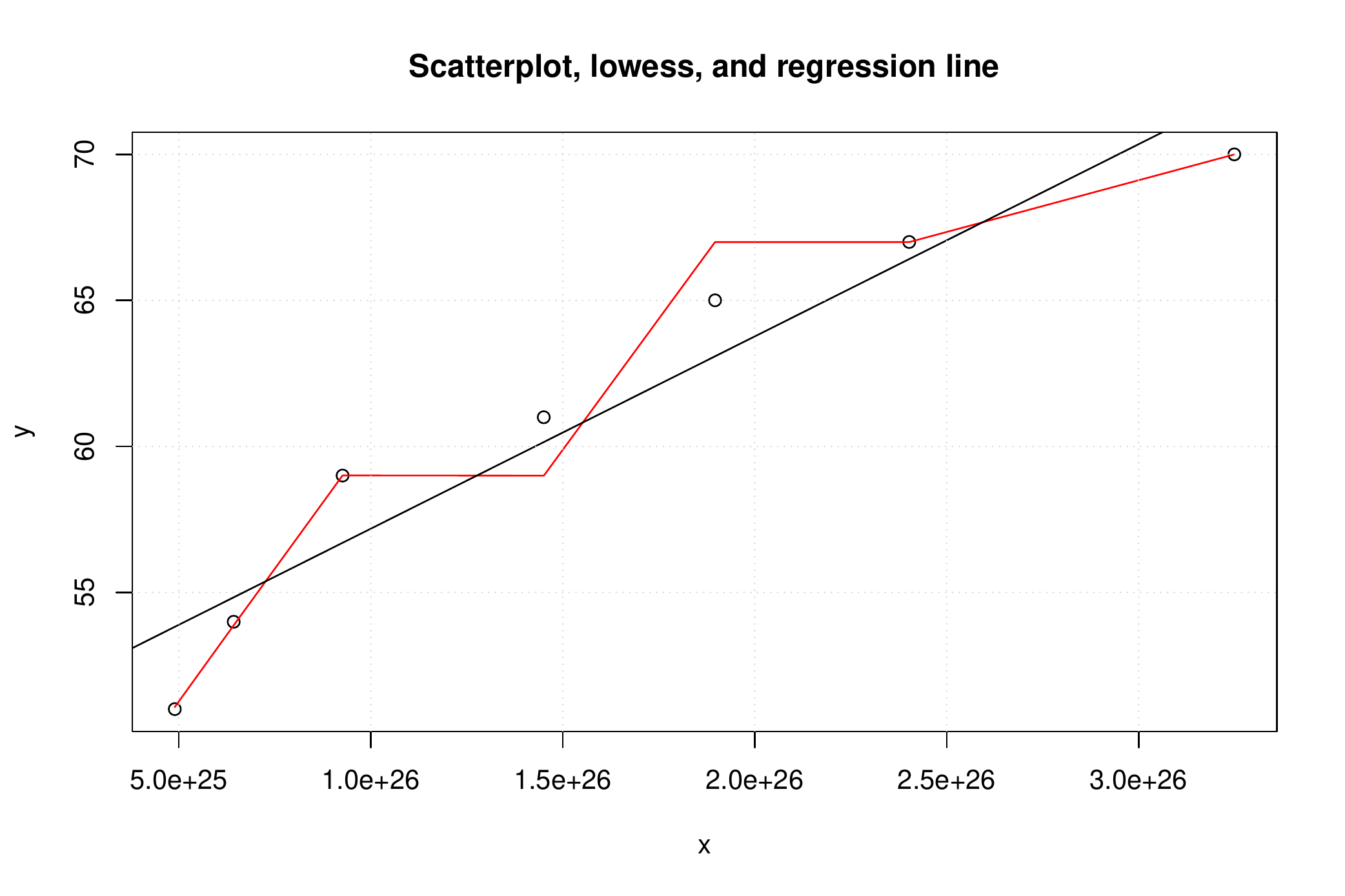}
	%
	%
	\caption{Regression line and Scatter Plot for the Data of Delhi from 2012--2018}
	\label{fig:regression}       
\end{figure}

The linear regression applied to the historical data of Delhi gives us an estimate of free-flow travel time and capacity. The figure \ref{fig:regression} gives the regression line for better understanding of the trend.
To understand the travel time, we study the traffic scenarios in different metro cities. We show for the metro city Delhi, the increment in travel time follow the BPR model as described in \eqref{eq:02}.

We deduce that the highly congested metro cities have a strong relationship between traffic volume and travel time over the years. 

For better insight, we see the drift in travel time by analysing the data of Delhi. We show that for Delhi, travel time follow the BPR trend as mentioned in \eqref{eq:02}. The cities that do not have a strong relationship between the volume of traffic and travel time are excluded. The reasons for excluding can be an error in data collection or the traffic volume not being the main factor for increase in travel time. We exclude these cities in analysis as the traffic volume does not facilitate in predicting the commute time. The exclusion  is done based on some statistical analysis. We measure the linear correlation between travel time $\mathcal{T}$ and the fourth-order of traffic volume $N^{4}$, and calculate their Pearson's correlation coefficient and two-tailed p-value. We filter out the metro areas where Pearson's correlation coefficient is larger than $0.5$, and p-value is smaller than $0.1$. On the basis of threshold, we predict that the analysis is applicable to the cities including Delhi, Bangalore, Mumbai and Kolkata and some other that are not included in the further analysis. The BPR model is then applied to Delhi and other metro cities. 
We see that in Delhi, travel time increases with  the increase in the number of vehicles, nicely following the BPR function.

We can observe a similar trend in the other mentioned metro cities using the BPR model, which gives us the estimated free-flow time and practical road capacity. 
Using the estimated free-flow travel time and road capacity, we calculate two important metrics for mentioned metro area:
\begin{itemize}
	\item Travel time ratio: the ratio of actual travel time versus free ﬂow travel time. A travel time ratio of $1.1$ means the current travel time is $1.1$ time (or $10\%$ higher) than the empty road travel time. 
	\item  Capacity ratio: the ratio of traffic volume versus road capacity. A capacity ratio of $1.1$ means there are $10\%$ more cars than the road capacity.
\end{itemize}
Travel time ratio and capacity ratio allows us to standardize each city, accounting for the fact that larger cities tend to have larger capacities and longer commutes, even when the roads are empty.

\begin{figure}[H]
	\centering
	\sidecaption
	\includegraphics[scale=.75]{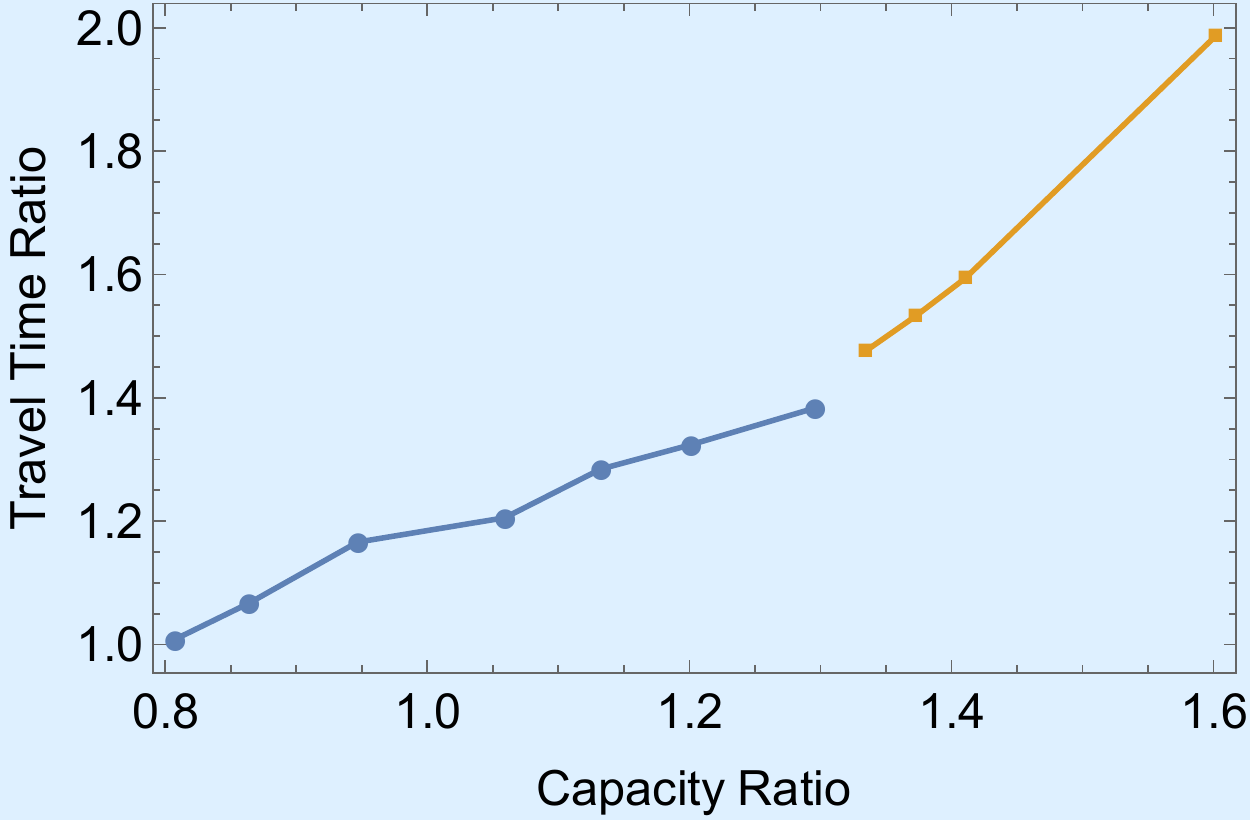}
	%
	%
	\caption{Travel time Ratio vs. Capacity Ratio for Delhi under different scenarios.}
	\label{fig:ratio}       
\end{figure}

Figure \ref{fig:ratio} shows the standardized relationship of travel time ratio and capacity ratio of Delhi (Blue dots are historical data and yellow dots are prediction for $25\%$, $50\%$, $75\%$ and $100\%$ switch to SOV). We see that the city follows the same growing curve when there is a decrease in the number of shared vehicles on the road. Essentially, what we are doing in the standardization, is looking at the rate travel time ratio increases with capacity ratio, then estimate where the city sits on the BPR curve. For Delhi, there is an increase in travel time ratio with an increase in capacity ratio. This indicates that the traffic volume of Delhi is way above road capacity. The same can be observed in Figure \ref{fig:ratio}.

\vspace{.35cm}
 \subsection{Prediction of Traffic Scenarios Post-Lockdown} 

%
\noindent For decades now, there has been a massive amount of pollutants in the atmosphere, which is a cause of major concern. The air quality has however enormously improved from hazardous to moderate during the lockdown imposed due to COVID-19.

A massive rise of pollutants is expected post-lockdown with increasing number of vehicles on roads with commuters switching to SOV.
A sudden drift in the number of vehicles will cause an adverse effect on the traffic system. Although, a massive decrement in the amount of pollutants in the atmosphere has been observed, which was a cause of major concern pre-lockdown. It has been successfully reduced, improving the air quality from hazardous to moderate.

\vspace{.35cm} 
According to the Air Quality Index, the unprecedented growth in the pollutants in the atmosphere has caused not only environmental and health issues but a heavy impact on the economy.

\vspace{.35cm} 
If we focus on the situation on the road traffic before lockdown, we see a high level of congestion in four megacities in India, namely Delhi, Mumbai, Bangalore and Kolkata where total congestion cost was estimated to be more than USD 22 billion per year. We can thus estimate the potential positive impact on ridesharing in each city.

\vspace{.35cm} 
Delhi and Mumbai have been two mega urban centres with relatively more developed modern public transport system. The congestion levels remained high during the pre-lockdown era due to high level of private vehicles in the city. With the burgeoning population and the growing prosperity of Delhi and Mumbai, the reliance on cars has increased, adding more pressure to the road network.

\vspace{.35cm} 
Bangalore and Kolkata have a relatively smaller population; these two cities have been more dependent on motorbikes and cars adding to huge ownership of private vehicles causing intense congestion. 

\vspace{.35cm} 
Government has urged the cities to maintain control over vehicle growth and encourages the use of public transport as key objectives for going forward. A combination of infrastructure improvement, addition of more mass transit, as well as efficient alternatives to vehicle ownership, will facilitate in curbing congestion.

\vspace{.35cm}
It can be easily predicted that if commuters quickly return to road, the transit users are more likely to switch to SOV. We have considered three scenarios namely when $25\%$, $50\%$, $75\%$ and $100\%$ of the carpool and transit users switch to SOV. The assumed scenarios are created to show a range of possibilities of the mode shift to SOV, which is highly likely post-lockdown in the country. It does not necessarily predict the actual mode shift. 

\begin{figure}[H]
	\centering
	\sidecaption
	\includegraphics[scale=.48]{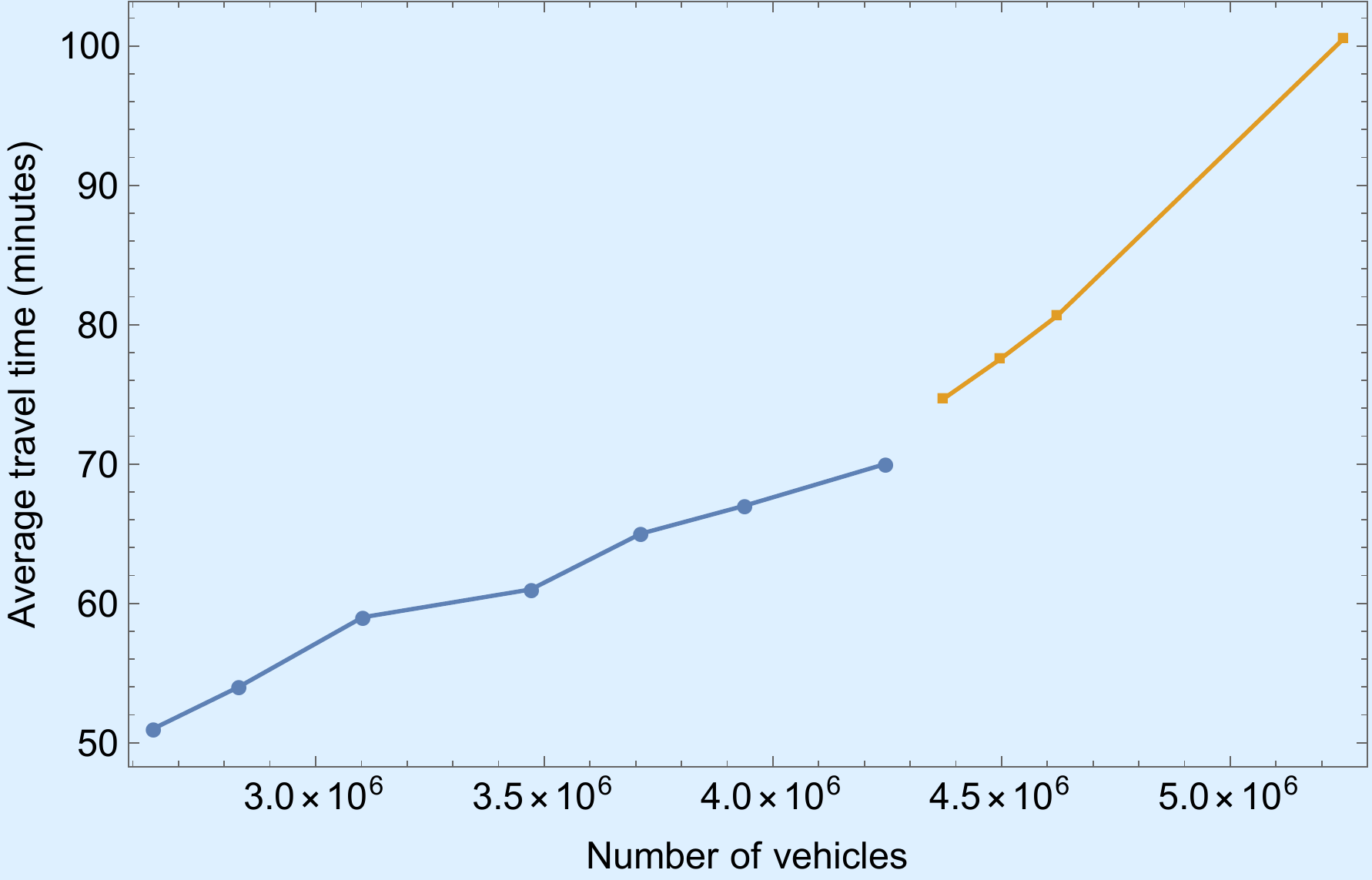}
	%
	%
	\caption{Example of BPR travel time prediction for Delhi. The blue points are the observed data, and the yellow points are the predictions when $25\%$, $50\%$, $75\%$ and $100\%$ of the carpool and
		transit commuters switch to SOV.}
	\label{fig:main1}       
\end{figure}

The predictions for Delhi under each of the four scenarios is shown in Figure \ref{fig:main1}.
The data plotted using the BPR model can be seen to be easily fitted along the curve. We can determine from figure \ref{fig:main1} that cities having more cars than the network can handle (for e.g. Delhi here) are sensitive to the changes in the number of vehicles on the road. A few minutes of increment in average travel time can make a huge impact on the traffic network, if multiplied by the number of road users in the city and this can result in thousands of additional hours time spent in traffic each day.
However, the rate of actual mode shift of transit and carpool users that will shift to personal vehicles or SOV depends on a multitude of factors. They may be specific of each traveller and each city, such as unemployment, cost of maintaining SOV, remote work etc. 

We are, however, not claiming that a specific percentage of users will switch to Single Occupied Vehicles.  The purpose of creation of the predicted scenarios is to identify the cities which are more sensitive to changes in the number of cars on the road post-lockdown in the country. 

\vspace{.25cm}
\section{India's Transport Growth Journey and its Effect on Energy and Environment}
\noindent If we look at the present situation in India, there is an urgent need to conserve energy and land, control pollution and ``greenhouse gas emissions'', and to alleviate poverty. 

Urban transport has been a major cause and a solution to combat these issues. The need of the hour is planned urban mobility solutions where all categories of road users are facing problems in commuting. 

If we look at the scenario before the pandemic brought the world to a standstill the increment in the number of vehicles on the road has caused problems to everybody, the pedestrians do not get a safe, conflict-free and obstruction-free path to walk. The cyclists have to fight for the space to cycle with fast moving motorized modes of transport, many a time risking their lives. This is due to over-crowding of vehicles reducing the area of parking, compelling people to park their vehicles out of the allotted area. 

The users of Shared Vehicles face long waiting periods, uncertainty in travel time and difficult conditions of travel. The movement of personal motorized modes of transport is slowed down by the slow-moving passenger and goods traffic and face significant delays at traffic signals and road junctions. Road users get restless leading to road rage, rash driving and accidents.

The use of desirable modes; walk, bicycle and PT is declining and the use of undesirable modes, i.e. car and 2-wheelers are growing. As a result, congestion is increasing, urban mobility as well as road safety are declining and, pollution, use of fossil fuel and accidents are rising everyday.

\subsection{Transport and Environment}
\noindent The second largest consumer of energy in India is the Transport Sector. The unprecedented and haphazard growth of the transport has been a major cause of concern, for it has not only increased pressure on the limited non-renewable energy resources but has considerably increased environmental pollution. 
 
Increasing car dependency in India, especially in the urban areas, is most visible in vehicular emissions which cause air pollution, noise pollution, and corresponding health effects. Increasing energy consumption, pollution, land intrusion and congestion are some of the areas that need urgent attention.

Transport planning is intrinsically linked to land use planning and both need to be developed together in a manner that serves the entire population and yet minimizes travel needs. An integrated master plan needs to internalize the features of sustainable urban transport. 

Rapid motorization of Indian cities has led to a public health crisis in the form of increased traffic injuries and fatalities, exposure to air and noise pollution, and decreased physical activity among many other adverse health and environmental impacts.

\vspace{.35cm}
\subsection{Health and Social Issues}
\begin{figure}[H]
	\centering
	\sidecaption
	\includegraphics[scale=2]{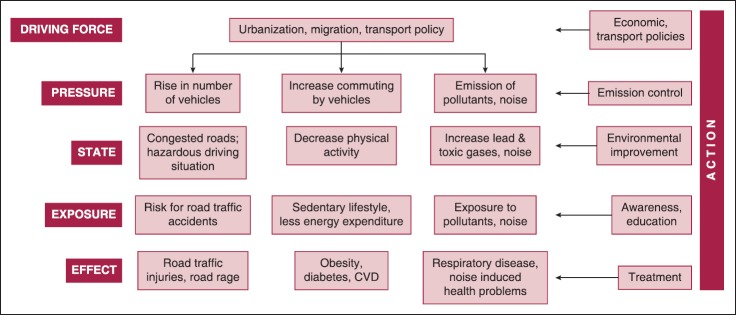}
	%
	%
	\caption{Effects of Road Transport on Growth. Source: ncbi.nlm.nih.gov/pmc/articles/PMC4746948/}
	\label{growth}       
\end{figure}

\noindent Adverse effects of increasingly car-based urban transport systems in India and other emerging economies are most visible on the local level. Vehicle emissions such as particulate matter, $NO_{x}$ or $SO_{x}$ pollute the atmosphere. Traffic noise act as a hindrance in conversation and disturbs the sleep. Road accidents pose further risk to the citizens' health, affecting especially vulnerable groups such as pedestrians, cyclists or children. 

The transport sector has contributed significantly to the emissions of toxic substances into the atmosphere. High demand for used automobiles might have sufficed and catered the need of the people, but it has made the situation worse for their health. Every vehicle on the road adds to the harmful substance in the atmosphere. 

Many surveys reveal that $45\%$ of the respondents claimed that they have transport-related diseases in the city. It is pertinent to note that $50\%$ of the respondents who claim that they have transport-related health challenges have eye problems, approximately $16\%$ has asthma and the same percentage has skin-burn diseases. In addition, $8\%$ and $5\%$ of the patients claim that they have upper respiratory tract infections and hypertension, respectively. Only $2\%$ of respondents indicated that they have hearing impairment resulting from unpleasant sounds emanating from indiscriminate use of horns by motorists and record players.

The increase in air pollution has very serious health implications. Poor air quality increases respiratory ailments like asthma and bronchitis, heightens the risk of life-threatening conditions like cancer, and burdens our health care system with substantial medical costs. Particulate matter is singlehandedly responsible for up to $30,000$ premature deaths each year.

It is impossible to reduce all environmental exposures to a level at which the risk to human health is zero. 

\vspace{.35cm}
\subsection{Personal Vehicles and their Impact}
\noindent Use of personal motorized vehicles and its significant contribution to air pollution, greenhouse gas emissions, and fossil fuel consumption are well accepted. The main reason for the increasing use of personal vehicles is the reluctance of the people to travel in Public Transports (PTs) for the inefficiency in maintaining the quality of the transport. Not until this situation was tackled that COVID-19 hit the country, making the situation even worse where even a greater population will now switch to a personal mode of commute, turning the whole situation gross.

Simultaneously, there is an urgent need to put a restraint on the use of personal vehicles. Government of India is already supporting measures such as traffic-calmed areas, pedestrianized areas, car limited zones, congestion pricing zones, no-emission zones, high parking charges, park \& ride facility and other economic instruments to control the indiscriminate use of personal motorized modes. However, these may not suffice in the current panic-like situation. 
 
Increase in the number of PT wouldn't suffice in such a situation of crisis. The pandemic is more likely to force people to choose Private Vehicle Ownership for the fear of infection, reducing their dependency on PTs. 

AQI calculates the air pollutants and particulate matter, nitrogen oxide, Sulphur dioxide, ozone, carbon monoxide etc. The ones that are of most concern is particulate matter with a diameter of $2.5$ and $10$ microns. The PT of these sizes cannot be removed and filtered from the body. 

It is obvious that low-levels of air pollution reached during the lockdown have considerably reduced the particulate matter in the atmosphere, thus reducing the deaths to approximately six lakhs. Due to the restricted activities during the period of lockdown, the particulate matter was reduced by $52\%$ nationwide. 

According to the Central government's System of Air Quality and Weather Forecasting and Research, all the sources of pollution have decreased during the lockdown period to a considerably lower level which has not been seen in the past four decades.

These changes are however temporary and once the lockdown is brought to closure, the vehicles will increase at an alarming rate due to a sudden switch to SOV or personal vehicles. A measure to save oneself from one deadly disease, i.e. COVID-19 would push them to the risk of another.

This situation needs dire and urgent attention to the public transport in India is not very developed and walkable and cycling environment needs to be promoted at a larger scale. With lakhs of travel trips a day, cities without adequate measures to tackle the situation will lock in an enormous amount of pollution and carbon. 

There is a rise of 7-14 time more pollutants contributed by car or two-wheelers than by buses in Delhi in every trip. There is a constant decline in the usage of the buses in Delhi which is expected to witness even a steeper decline post lockdown. 

\subsection{Measures to Curb the Traffic Upsurge} 
\noindent With the unlocking of the cities, there will be steep inclination in the people switching to private mode of transport. The spread of the infectious disease can be easily contracted in close proximity of the infected person therefore, people will keep off from enclosed spaces especially the shared mode of transport.  
 
Confined and crowded environment result in easy contraction of the disease which might result in an outbreak. The restarting of the public transport with reduced crowding is an issue of paramount importance. 
 
The safety measures should be incorporated in such a manner to protect commuters and on-board staff. Proper and regular sanitization of public transport with proper distancing might serve as a boon. 

Several other ways are being discussed by the government keeping in view the welfare of the society.

\section*{Conclusion}
\noindent After performing the experiment with the help of the given data, it is observed that the increase in the average travel time is only by few minutes. This may appear as a minimal change but the delay takes place in each trip. To understand the actual time-lapse if it is multiplied by the total number of travellers in the city, even a minute of change will result in thousands of additional hours in the traffic.

\vspace{.25cm}
The data on road commute mode and travel time in different cities has been accumulated from the verified sources such as State-Resource-Centre, NITI Aayog and Transport Sector of India.  

\vspace{.25cm}
A BPR model is further used to establish a relation between the average travel time to the estimated number of commuters travelling by car. The resulting travel time is thus deduced using the BPR model by estimating the number of cars on the road wherein the transit and carpool users switch to single-occupancy vehicles.

\vspace{.25cm}
Therefore, it can be concluded that
\begin{itemize}
	\item When the number of vehicles is more than the capacity of the road. The increasing cars are inimical to everyone's commute. This is calculated and analyzed using the BPR model.
	
	\vspace{.15cm}
	\item COVID-19 brought serious implication in the mode shift of traffic. The imposition of lockdown throughout the country resulted in deserted roads thereby decreasing the pollution. With the reopening traffic will eventually spring back. If transit ridership does not return, travel time will increase, sometimes dramatically.
	
		\vspace{.15cm}
	\item A possible increase in travel time per trip is of 5-20 minutes in high-transit cities, which add up to several hundreds of thousands of hours of travel time each day.
	
		\vspace{.15cm}
	\item These discordant and erratic increases are avoidable, if transit ridership resumes in accordance with car traffic.
\end{itemize}
\eject%
%





%
%



%
\begin{acknowledgement}
The authors are thankful to Government of India and their respective bodies for putting up important data related to urban transport which was the main source of data for the analysis.
\end{acknowledgement}

\end{document}